\documentclass[10pt,conference]{IEEEtran}
\IEEEoverridecommandlockouts
\usepackage{cite}
\usepackage{amsmath,amssymb,amsfonts}
\usepackage{algorithmic}
\usepackage{graphicx}
\usepackage{textcomp}
\usepackage{xcolor}

\usepackage{bm}
\usepackage{customcommands}
\usepackage{siunitx}
\usepackage{booktabs}
\ifCLASSOPTIONcompsoc
\usepackage[caption=false,font=normalsize,labelfon
t=sf,textfont=sf]{subfig}
\else
\usepackage[caption=false,font=footnotesize]{subfi
g}
\fi

\def\BibTeX{{\rm B\kern-.05em{\sc i\kern-.025em b}\kern-.08em
    T\kern-.1667em\lower.7ex\hbox{E}\kern-.125emX}}

\begin{document}

\bstctlcite{IEEEexample:BSTcontrol}

\title{Solving 1D Poisson problem with a Variational Quantum Linear Solver
}

\author{\IEEEauthorblockN{Giorgio Tosti Balducci}
\IEEEauthorblockA{\textit{Aerospace Structures and Materials} \\
\textit{TU Delft}\\
Delft, The Netherlands \\
g.b.l.tostibalducci@tudelft.nl}
\and
\IEEEauthorblockN{Boyang Chen}
\IEEEauthorblockA{\textit{Aerospace Structures and Materials} \\
\textit{TU Delft}\\
Delft, The Netherlands \\
b.chen-2@tudelft.nl}
\and
\IEEEauthorblockN{Matthias M\"{o}ller}
\IEEEauthorblockA{\textit{Applied Mathematics} \\
\textit{TU Delft}\\
Delft, The Netherlands \\
m.moller@tudelft.nl}
\and
\IEEEauthorblockN{Roeland De Breuker}
\IEEEauthorblockA{\textit{Aerospace Structures and Materials} \\
\textit{TU Delft}\\
Delft, The Netherlands \\
r.debreuker@tudelft.nl}
}

\maketitle

\begin{abstract}
Different hybrid quantum-classical algorithms have recently been developed as a near-term way to solve linear systems of equations on quantum devices. However, the focus has so far been mostly on the methods, rather than the problems that they need to tackle. In fact, these algorithms have been run on real hardware only for problems in quantum physics, such as Hamiltonians of a few qubits systems. These problems are particularly favorable for quantum hardware, since their matrices are the sum of just a few unitary terms and since only shallow quantum circuits are required to estimate the cost function. However, for many interesting problems in linear algebra, it appears far less trivial to find an efficient decomposition and to trade it off with the depth of the cost quantum circuits. A first simple yet interesting instance to consider are tridiagonal systems of equations. These arise, for instance, in the discretization of one-dimensional finite element analyses.

This work presents a method to solve a class of tridiagonal systems of equations with the variational quantum linear solver (VQLS), a recently proposed variational hybrid algorithm for solving linear systems. In particular, we present a new decomposition for this class of matrices based on both Pauli strings and multi--qubit gates, resulting in less terms than those obtained by just using Pauli gates. Based on this decomposition, we discuss the tradeoff between the number of terms and the near-term implementability of the quantum circuits. Furthermore, we present the first simulated and real-hardware results obtained by solving tridiagonal linear systems with VQLS, using the decomposition proposed.
\end{abstract}


\section{Introduction}
Current quantum computers are a tool in search for applications. Even though several quantum algorithms have deep practical implications and offer up to exponential speed-ups, their promises are so far purely theoretical. Good examples of this are Shor's algorithm for period finding \cite{Shor1999} and Grover's for database search \cite{Grover1996}.

The reason why no implementations at useful scales exist for this algorithms is that quantum hardware is still far from its maturity. In fact, we are  in the so-called Noisy Intermediate Scale Quantum (NISQ) era \cite{Preskill2018}, in which quantum computers have 50 to 100 physical qubits, that are noisy and mostly not error-corrected.

As a consequence, one of today's challenges is to find problem-method combinations that can run on NISQ devices with limited error. There are ultimately two ways to go. In the first one a problem is sought for that is classically hard but quantum efficient, regardless of the practical interest in the problem itself \cite{Arute2019}. These are the applications aimed at proving quantum supremacy. A second case is when a NISQ computer is used to solve a relevant problem. Here however, the algorithms are not the ones promising a speed-up on fully connected and error-corrected quantum hardware, but others that take into account the limitations of today's machines. Proabably the most prominent class of these methods are the so-called Variational Quantum Algorithms (VQAs) \cite{McClean2016}. These already found applications in different disciplines, such as quantum chemistry \cite{Peruzzo2014, OMalley2016, Jones2019} and quantum machine learning \cite{Schuld2019,Chen2020,Romero2020}.

Recently VQAs were also developed for solving linear system of equations \cite{huang2019nearterm,xu2019variational,Bravoprieto2020variational,An2020quantum,Wu2021}, a problem encountered in many different scientific and engineering disciplines. Unlike the seminal work of Harrow, Hassiddim and Lloyd on solving linear systems on an ideal quantum device \cite{Harrow2009}, these algorithms do not promise an exponential speed-up. However, they were applied successfully to problems of dimensions $2^{300}\times 2^{300}$ on a simulator \cite{huang2019nearterm} and $1024\times 1024$ on a quantum computer \cite{Bravoprieto2020variational}. To the best of the authors' knowledge, the latter is the largest linear systems ever to be solved on quantum hardware.

However, the linear systems in previous works, which were solved on quantum hardware with a variational linear solver \cite{Bravoprieto2020variational} can be thought of as technology demonstrators. In fact, in these instances, the system's matrix is always chosen as a Hamiltonian acting on few qubits, that can be efficiently encoded on a quantum computer. The question yet to be answered is whether variational linear solvers can efficiently solve problems that stem from real-world applications and that are not specifically tuned to the hardware. A simple, yet interesting starting point are tridiagonal linear systems, which derive, for instance, from the discretization of the Poisson problem in one dimension \cite{Quarteroni2014}.

In this work, we adopt a VQA for linear systems, namely the Variational Quantum Linear Solver (VQLS) \cite{Bravoprieto2020variational} and focus on the problem of decomposing the Poisson 1D discrete matrix in unitary terms. This decomposition is an important step, as the algorithm's efficiency depends on the number of unitary components, since this affects the amount of quantum circuits to evaluate. To this end, we compare the naive decomposition in Pauli strings with a novel approach, that includes multi-qubit gates. We show that the new decomposition almost halves the number of unitary terms, but also that these correspond to deeper, thus less near-term quantum circuits. Finally, we present and discuss the results obtained by solving small tridiagonal linear systems both on a simulator and on quantum hardware.

We ultimately remark that this work is not intended as isolated. In our view, this will be the first step before extending the application of VQAs to other linear systems, which derive from the numerical discretization of partial differential equations.

\section{Background}
\subsection{The Quantum Linear System Problem}
Let $A$ be a square complex $N \times N$ matrix and $\bm{b}$ a complex $N$-dimensional vector. The aim is to find the complex vector $\bm{x}$ such that
\begin{equation}
    A \bm{x} = \bm{b}.
    \label{eq:LSP}
\end{equation}

In general, \eqref{eq:LSP} is not in a form suitable for quantum computation, since, $\bm{x}$ and $\bm{b}$ do not necessarily have unit norm and thus are not valid quantum states. Nevertheless, an analogous problem can be solved, that is
\begin{equation}
    A \ket{x} = \ket{b}, 
    \label{eq:QLSP}
\end{equation}
where $\ket{x} = \bm{x}/\norm{x}$ and $\ket{b} = \bm{b}/\norm{b}$. Eq. \eqref{eq:QLSP} is known as the \emph{Quantum Linear System Problem} (QLSP).

Solving \eqref{eq:QLSP} on a quantum computer requires $A$ and $\ket{b}$ to be encoded in terms of unitary gates and quantum states. This means that the matrix $A$ must be decomposed as
\begin{equation}
    A = \sum_{l=1}^L c_l A_l,
    \label{eq:mat_decomp_generic}
\end{equation}
where $c_l \in \mathbb{C}$ and $A_l$ are unitary $N\times N$ complex matrices. Also, the state $\ket{b}$ needs to be prepared as
\begin{equation}
    \ket{b} = B \ket{0} = \sum_{m=1}^M c_m B_m \ket{0},
    \label{eq:rhs_prep_generic}
\end{equation}
with $c_m$ and $B_m$ being, respectively, complex scalars and unitary matrices.

As explained later in the paper, the problem of reducing the number of unitary components of $A$ and $B$ is crucial for the success and the efficiency of a variational linear solver.

\subsection{The VQLS algorithm}
The Variational Quantum Linear Solver \cite{Bravoprieto2020variational} is an algorithm that allows to find an approximate solution $\ket{x_{f}}$ to the generic QLSP.

In essence, it consists in minimizing a cost function that represents the distance between the states $\ket{\psi} = A\ket{x}$ and $\ket{b}$. The optimization parameters $\bm{\theta}$ determine the tentative solution $\ket{x}$ at every iteration as
\begin{equation}
    \ket{x(\bm{\theta})} = V(\bm{\theta})\ket{0},
    \label{eq:ansatz_eqn_generic}
\end{equation}
where $V(\bm{\theta})$ is a parametrized quantum circuit known as \emph{ansatz}. Analogously to all other variational quantum algorithms, VQLS is hybrid, since it runs on both a quantum and a classical computer operating in loop. While the quantum computer evaluates the unitary circuits that represent the cost function, the classical computer postprocesses the results of the quantum part and updates the ansatz parameters with an optimization routine.

In what follows, the different parts of VQLS will be discussed in more detail.
\subsubsection{Cost function}
a suitable cost function is the so-called \emph{normalized global cost} \cite{Bravoprieto2020variational}
\begin{equation}
    C_G = 1 - |\brakettwo{b}{\Psi}|^2,
    \label{eq:cost_global_norm}
\end{equation}
where $\ket{\Psi}=\ket{\psi}/\brakettwo{\psi}{\psi}$. This cost function vanishes when $|\brakettwo{b}{\Psi}| = \num{1}$, that is at the solution of \eqref{eq:QLSP}. It should be noticed that minimizing \eqref{eq:cost_global_norm} is equivalent to finding the ground state of the Hamiltonian
\begin{equation}
    H_G = A^\dagger (1 - \ket{b}\bra{b}) A ,
    \label{eq:hamilt_global}
\end{equation}
since $\bra{x} H_G \ket{x} = 0$ if $\ket{x} = A^{-1}\ket{b}$,

What motivates the use of a quantum computer in VQLS is the hardness of classically evaluating a cost function such as \eqref{eq:cost_global_norm}. In fact, if $n$ is the number of qubits on which the gates in the cost's expression act, the evaluation of \eqref{eq:cost_global_norm} to within precision $\delta = 1/\text{poly}(n)$ is a DQC1-hard problem \cite{Bravoprieto2020variational}. We remind that the Deterministic Quantum Computing with 1 Clean Qubit (DQC1) complexity class includes all those probems that can be solved in the 1-clean-qubit model of computation in polynomial time \cite{Knill1998} and that it is believed impossible to efficiently solve DQC1 problems with classical logic \cite{Morimae2017,Fuji2018}.

An important issue with variational algorithms concerns the number of quantum circuits to solve for obtaining the cost value. The cost in \eqref{eq:cost_global_norm} is estimated by computing $|\brakettwo{b}{\Psi}|^2$, for which we need to evaluate separetely $\brakettwo{\psi}{\psi}$ and $|\brakettwo{b}{\psi}|^2$. These terms can be written by replacing $A$ with the expansion from \eqref{eq:mat_decomp_generic}. For the sake of this explanation $B$ is assumed unitary and can therefore be implemented directly as a quantum circuit. The cost terms are therefore
\begin{equation}
    \brakettwo{\psi}{\psi} = \sum_{l=1}^L\sum_{l^\prime=1}^L c_l^* c_{l^\prime} \bra{x} A_l^\dagger A_{l^\prime} \ket{x},
    \label{eq:psipsi_with_A_expanded}
\end{equation}

\begin{equation}
    |\brakettwo{b}{\psi}|^2 = \sum_{l=1}^L\sum_{l^\prime=1}^L c_l^* c_{l^\prime} \bra{x} A_l^\dagger B \ket{0} \bra{0} B A_{l^\prime}\ket{x}.
    \label{eq:psibsq_with_A_expanded}
\end{equation}

It is easy to see that \eqref{eq:psipsi_with_A_expanded} and \eqref{eq:psibsq_with_A_expanded} both require to compute the expecation values of $O(L^2)$ unitaries, corresponding to as many quantum circuits evaluations. Therefore, it is essential in VQLS to decompose $A$ in as few unitary terms as possible.

A Hadamard Test circuit can be used to actually compute the expectation values in \eqref{eq:psipsi_with_A_expanded} and \eqref{eq:psibsq_with_A_expanded}. The resulting circuits may however be too deep, and the results could be affected by noise due to decoherence. This is especially a problem for the $\bra{x} A_l^\dagger B \ket{0}$ terms, where $V(\bm{\theta})$, $A_l$ and $B$ must all be controlled. Expedients such as the Hadamard-Overlap Test \cite{Bravoprieto2020variational} can partially solve this issue by removing the controlled $V(\bm{\theta})$ and controlled $B$, while doubling the number of qubits.

\subsubsection{Ansatz}
the ansatz circuit is a fundamental component of every variational algorithm.

In literature, different ansatz types exist, but a main distinction is between the so-called `hardware-efficient' ansatze \cite{Kandala2017} and the Quantum Alternating Operator Ansatze (QAOA) \cite{Hadfield2019}. The first ones are built by keeping the characteristics of quantum hardware in mind, such as connectivity and quality of the qubits. However, these ansatze are problem-agnostic, because they do not relate to the input of the problem. Oppositely, the QAOA is inspired by the adiabatic principle and then needs a final Hamiltonian that is aware of the original problem. Moreover QAOA is universal \cite{lloyd2018quantum}, since it can represent any quantum state once enough resources are available. In general, however, this class of ansatze does not take hardware feasibility into account by default. 

In practice, both the ability of the ansatz to represent the final solution and its implementation on hardware are important aspects. However, today's hardware has limited connectivity and qubit coherence times, thus imposing a choice on the type of ansatz. In particular, QAOAs should be used for problems of small dimensions, for which the solution should be accurate. On the other hand, increasing the problem's dimensions may only be achieved in NISQ with a Hardware-Efficient Ansatz, possibly at the cost of giving up some of the solution's accuracy.

\subsubsection{Optimization routine}
VQAs use a classical optimizer to find the global minimum of the cost function.

The matter of choosing an optimizer is in itself not trivial and very problem dependent. In general the main distinction is between gradient-based and gradient-free methods. The first ones include algorithms such as stochastic gradient descent \cite{Sweke2020} and BFGS \cite{Fletcher1987} and they can quickly lead to the minimum if the cost function has enough regularity. However, the noisy nature of the VQA's cost function and the possibility of barren plateaus may prevent gradient-based methods from converging. On the other hand, gradient-free methods have usually more noise resilience, which improves the chances to converge. Except VQA-specific methods, often classical optimizers such as Nelder-Mead \cite{Nelder1965} and COBYLA \cite{Powell1994} are chosen. However, even for these algorithms, barren plateaus in the cost function may prevent the optimizer from finding the cost's minimum.

\section{Decomposition of a tridiagonal matrix}
This work deals with solving QLSPs, where the matrix $A$ is tridiagonal and has constant coefficients. In particular, $A$ has the form
\begin{equation}
    \begin{bmatrix}
        \alpha  & \beta  &        &         &        &       &       \\
        \beta   &\ddots  & \ddots &         &        &       &       \\
                & \ddots &\ddots  & \ddots  &        &       &       \\
                &        & \beta  & \alpha  & \beta  &       &       \\
                &        &        &  \ddots & \ddots & \ddots&       \\
                &        &        &         & \ddots &\ddots & \beta \\
                &        &        &         &        &\beta  & \alpha
    \end{bmatrix},
    \label{eq:tridiag_generic}
\end{equation}
where $\alpha, \, \beta \in \mathbb{R}$.

As previously discussed, variational linear solvers require to decompose the matrix $A$ into unitary terms, according to \eqref{eq:mat_decomp_generic}. The rest of this section presents two ways to do so. The first one is the decomposition in tensor products of Pauli operators, also known as \emph{Pauli strings}, while the second one is specific to matrices of class \eqref{eq:tridiag_generic} and uses also multi-qubit gates.

\subsection{Naive Pauli decomposition}
Generally speaking, every $N\times N$ Hermitian matrix can be expressed as a linear combination of $N^2$ \emph{Pauli strings} \cite{Nielsen2010}, where each Pauli string acts on $n = \log_2(N)$ qubits. Since tridiagonal matrices of the form \eqref{eq:tridiag_generic} are symmetric, thus Hermitian, it holds that
\begin{equation}
    A = \sum_{i=1}^{N^2} c_i P_i,
    \label{eq:pauli_decomp}
\end{equation}
where $P_i$ is the i\textsuperscript{th} Pauli string. These have the form
\begin{equation}
    P_i = \sigma_{\alpha_1}^i \otimes \sigma_{\alpha_2}^i \otimes \dots \otimes \sigma_{\alpha_n}^i,
    \label{eq:pauli_string}
\end{equation}
where $\sigma_{\alpha_j}^i \in \{X,\, Y, \, Z, \, I\}$ are the Pauli operators. 

Since all Pauli strings are unitary, \eqref{eq:pauli_decomp} is a valid decomposition and can be used as input to VQLS. Also, tridiagonal matrices of the form \eqref{eq:pauli_decomp} have only $N$ of the $N^2$ terms with non-zero coefficients \cite{IEEEexample:Cappanera2021}. For instance, for a $4\times 4$ case, the Pauli decomposition is
\begin{equation}
    A = \alpha\, I_1 I_0 + \beta\, I_1 X_0 + \frac{\beta}{2}\, X_1 X_0 + \frac{\beta}{2}\, Y_1 Y_0,
    \label{eq:pauli_decomp_4x4}
\end{equation}
where we use the notation
\begin{equation}
    \sigma_{\gamma, 1} \, \sigma_{\delta, 0} = \sigma_{\gamma} \otimes \sigma_{\delta}.
    \label{eq:pauli_tensor_notation}
\end{equation}

\subsection{Decomposition with multi-qubit gates}
The set of Pauli strings is not the only option to decompose \eqref{eq:tridiag_generic}. One class of quantum gates that cannot be replicated by any single Pauli string  are multi-qubit gates, such as SWAP or controlled gates. Our novel decomposition makes use of exactly these multi-qubit gates to reduce \eqref{eq:tridiag_generic} to unitary terms.

In particular, the use of SWAP and SWAP-like gates, which we name \emph{center-switch} gates, almost halves the number of terms with respect to the Pauli decomposition. The intuition behind it lays in the matrix strucure of these operators. In fact, every Pauli string is expressed by a block-diagonal matrix, where the blocks can be either on the main diagonal or on the anti-diagonal. This constraint on the structure makes it impossible for a single Pauli gate to reproduce the center off-diagonal terms of \eqref{eq:tridiag_generic}, such as those on the left of Tab \ref{tab:critical_term_cfr}. On the other hand, the SWAP and center-switch gates have a matrix form with exactly the center off-diagonal entries of a tridiagonal matrix. 

In this section, a simple $\num{4}\times \num{4}$ matrix will first be considered. This is a simple, yet clear example of the importance of SWAP-like terms. Later on, the generic $\num{2}^n\times \num{2}^n$ case will be examined, which extends the idea and allows to introduce the center-switch gate.

\begin{table*}[t]
    \renewcommand{\arraystretch}{1.3}
    \caption{Two different ways of reproducing the off-diagonal entries in the center of a tridiagonal matrix. For a $4\times 4$ matrix, two Pauli strings are required, in order to compensate for the unnecessary terms in the anti-diagonal blocks. On the other hand, the SWAP gate alone includes the center off-diagonal entries.}
    \label{tab:critical_term_cfr}
    \centering
    \begin{tabular}{ c c c }
        \toprule
         & $XX + YY$ & SWAP\\
        \midrule\\
        \addlinespace[-2ex]
        $\begin{bmatrix}  0 & 0 & 0 & 0 \\ 0 & 0 & 1 & 0 \\ 0 & 1 & 0 & 0 \\ 0 & 0 & 0 & 0 \end{bmatrix}$ &
        $\begin{bmatrix}  0 & 0 & 0 & 1 \\ 0 & 0 & 1 & 0 \\ 0 & 1 & 0 & 0 \\ 1 & 0 & 0 & 0 \end{bmatrix} + \begin{bmatrix}  0 & 0 & 0 & -1 \\ 0 & 0 & 1 & 0 \\ 0 & 1 & 0 & 0 \\ -1 & 0 & 0 & 0 \end{bmatrix}$&
        $\begin{bmatrix}  1 & 0 & 0 & 0 \\ 0 & 0 & 1 & 0 \\ 0 & 1 & 0 & 0 \\ 0 & 0 & 0 & 1 \end{bmatrix}$\\
        \bottomrule
    \end{tabular}
\end{table*}

\subsubsection{$\num{4}\times \num{4}$}
Tab. \ref{tab:critical_term_cfr} shows the center off-diagonal entries of a $4\times 4$ tridiagonal matrix, together with the terms of two decompositions that replicate them. It should be clarified, that only the matrix structure is important in this discussion. Thus, the coefficients, as well as $\alpha$ and $\beta$ are chosen equal to 1, regardless of whether the terms in Tab. \ref{tab:critical_term_cfr} balance out or not.

It can be seen that, while two Pauli terms are needed to reproduce the structure on the left hand side of Tab \ref{tab:critical_term_cfr}, the SWAP gate alone has the required off-diagonal entries at the center of the tridiagonal matrix. When the linear system is of such small dimensions, however, introducing a multi-qubit gate does not reduce the overall number of terms to decompose \eqref{eq:mat_decomp_generic}. In fact, one still needs to compensate for the unnecessary diagonal terms of the SWAP, by introducing the $Z_1 Z_0$ Pauli string. Indeed, both the Pauli decomposition and the proposed one require four terms, as seen in Tab. \ref{tab:decompositions_cfr}. However, as remarked later, the diagonal terms can be corrected with just $n$ unitary terms, whereas $2^{n-1}$ Pauli strings are necessary in order to reproduce the center off-diagonal terms. This certainly provides a saving for more interesting problem sizes.

\subsubsection{$\num{2}^n\times \num{2}^n$, $n > 2$}
to reproduce the center off-diagonal terms of a $\num{2}^n\times \num{2}^n$  tridiagonal matrix with $n > 2$, a new gate type is needed. The basic requirements are for the gate to be unitary and to behave similarly to SWAP for the $4\times 4$ case. By looking at the $8\times 8$ case, it is easy to see that the gate in \eqref{eq:center_switch}, that we refer to as \emph{center-switch}, satisfies both previous requirements
\begin{equation}
    \text{CS} = 
    \begin{bmatrix}
        1&&&&&&& \\
        &1&&&&&& \\
        &&1&&&&& \\
        &&&&1&&& \\
        &&&1&&&& \\
        &&&&&1&& \\
        &&&&&&1& \\
        &&&&&&&1
    \end{bmatrix}.
    \label{eq:center_switch}
\end{equation}

In order to have a hardware implementation, CS needs to be reduced to more standard gates. In particular, CS can be expressed as a sequence of Toffoli gates, by using basic permutation theory.
To start with, it should be recalled that a \emph{transposition} is simply a permutation between two variables, i.e.
\begin{equation}
    (a,\, b): \quad a\rightarrow b \quad \text{and} \quad b\rightarrow a .
    \label{eq:transposition}
\end{equation}
By introducing a third variable $p$, $(a,\, b)$ can be expanded as a product of three transpositions, thus
\begin{equation}
    (a,\, b) = (a,\, p)\: (p,\, b)\: (a,\, p).
    \label{eq:transposition_expansion}
\end{equation}
Eq. \eqref{eq:transposition_expansion} simply expresses the computer progamming practice of swapping two variables by means of an intermediate one.

In particular, considering transpositions between bitsrings, then
\begin{equation}
    \text{CS} = (\text{011}, \text{100}),
    \label{eq:center_switch_transp}
\end{equation}
where bitstrings are written in little-endian notation. Furthermore, the transposition in \eqref{eq:center_switch_transp} can be expanded twice by using \eqref{eq:transposition_expansion} and two intermediate bitstrings, e.g., $p_1 = \text{001}$ and $p_2 = \text{000}$, such that
\begin{align}
    \begin{split}
        \label{eq:center_switch_transp_exp}
        \text{CS} &= (011,\, 100)
            = (011,\, 001)(001,\, 100)(011,\, 001)\\
            &= (011,\, 001)(001,\, 000)(000,\, 100)(001,\, 000)(011,\, 001).
    \end{split}
\end{align}

\begin{figure}[tbp]
    \centerline{\includegraphics[scale=1.5]{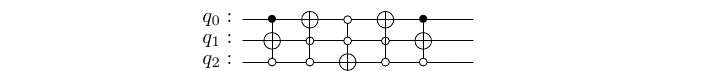}}
    \caption{Center-switch gate corresponding to the $(011,\, 100)$ permutation. Qubits are ordered from least to most significant, going from $q_0$ to $q_2$. Each one of the gates corresponds to a Toffoli gate, i.e. gates having two control qubits and one target. An empty dot means that the corresponding qubit must be in state 0 to flip the target qubit, and the analogous holds for a full dot and state 1. This figure was produced using the IBM Qiskit library \cite{Qiskit}.}
    \label{fig:center_switch}
\end{figure}

The gates corresponding to the transposition in the last line of \eqref{eq:center_switch_transp_exp} are Toffoli gates. For instance the $(011,\, 001)$ term corresponds to a Toffoli gate whose control qubits are the most and least significant ones and whose target qubit is the middle one. The middle qubit is flipped when the first control is in state 0 and the second one is in state 1. Fig \ref{fig:center_switch} shows the circuit implementing \eqref{eq:center_switch_transp_exp}.

The concept of the center-switch gate can be extended to the case of $n$ qubits and $2^n \times 2^n$ matrices. In fact, CS can be generalized as
\begin{equation}
    \text{CS}^{(n-2)} = (011\dots 1, \, 100 \dots 0),
    \label{eq:center_switch_gen}
\end{equation} 
where the bitstrings are both $n$ bits long. Following the same reasoning as in the $8\times 8$ case, each $\text{CS}^{(n-2)}$ can be decomposed as the product of $2n - 1$ multi-controlled gates, each with $n - 1$ control qubits.

The set of Pauli gates and their tensor products, together with the SWAP and $\text{CS}^{(n-2)}$ gates are all the unitary terms required to decompose a tridiagonal $2^n\times 2^n$ matrix of type \eqref{eq:tridiag_generic}.

\subsection{Comparison}
Tab. \ref{tab:decompositions_cfr} shows the unitary components for both the Pauli decomposition and the proposed one for dimensions up to $\num{16} \times \num{16}$. It is noticed how the introduction of SWAP and CS gates effectively avoids to use $N/2$ tensor products of $X$ and $Y$ gates that are necessary in the Pauli decomposition. In this way, the objective of reducing the number of unitary components of the tridiagonal matrix is achieved. Nevertheless, SWAP and CS introduce superfluous diagonal entries that need to be compensated for. This is the reason why the right hand side column of Tab. \ref{tab:decompositions_cfr} shows tensor product terms of Pauli $Z$ gates. However, this correction is not nearly as expensive as the one for the center off-diagonal components, since it only requires $n$ additional unitaries. The final count is then $2^{(n-1)} + n$ components for the multi-qubit gates decomposition, against $2^n$ terms for the Pauli decomposition.

This saving, however, is obtained by introducing gates that are more complex than Pauli strings. Since these gates enter the cost function's expression, they ultimately affect the quantum circuits to compute at every iteration of VQLS. Eventually, the cost circuits in \eqref{eq:psipsi_with_A_expanded} and \eqref{eq:psibsq_with_A_expanded} are deeper, thus more noise-sensitive when implementing SWAP and CS gates, rather than only Pauli strings. As an example, the cost circuits required to solve a $4\times 4$ linear system can be compared.
Keeping the same $B$, the same ansatz,  and the same reference hardware, the maximum circuit depths in the Pauli and multi-qubit gates cases are in a 51/99 ratio. This suggests that, although being more efficient, the new approach is not as near term as the standard Pauli decomposition technique.

\renewcommand{\arraystretch}{1.5}
\begin{table*}[htbp]
    \caption{Comparison between two different decompositions of the $A$ matrix. The Pauli decomposition requires $N = 2^n$ Pauli strings, while, using multi-qubit gates, only  $2^{n-1}+n$ unitary terms are needed.}
    \begin{center}
        \begin{tabular}{c c c c c}
        \toprule
        & $n$ & $N$ & Pauli decomposition & Multi-qubit decomposition \\
        \midrule
        & 2 & 4 & $I_1 I_0$, $I_1 X_0$, $X_1 X_0$, $Y_1 Y_0$& $\text{SWAP}_{(1-0)}$, $I_1 I_0$, $Z_1 Z_0$, $I_1 X_0$\\
        & 3 & 8 & \begin{tabular}{c} $I_2 I_1 I_0$, $I_2 I_1 X_0$, $I_2 X_1 X_0$, $I_2 Y_1 Y_0$, \\ $X_2 X_1 X_0$, $X_2 Y_1 Y_0$, $Y_2 X_1 Y_0$, $Y_2 Y_1 X_0$ \end{tabular}  & \begin{tabular}{c} $I_2 \text{SWAP}_{(1-0)}$, $I_2 I_1 I_0$, $I_2 Z_1 Z_0$, $I_2 I_1 X_0$\\ $\text{CS}_{(2-0)}$, $Z_2I_1Z_0$, $Z_2Z_1I_0$ \end{tabular}\\
        & 4 & 16 & \begin{tabular}{c} $I_3 I_2 I_1 I_0$, $I_3 I_2 I_1 X_0$, $I_3 I_2 X_1 X_0$, $I_3 I_2 Y_1 Y_0$, \\ $I_3 X_2 X_1 X_0$, $I_3 X_2 Y_1 Y_0$, $I_3 Y_2 X_1 Y_0$, $I_3 Y_2 Y_1 X_0$ \\ $X_3 X_2 X_1 X_0$, $X_3 X_2 Y_1 Y_0$, $X_3 Y_2 X_1 Y_0$, $X_3 Y_2 Y_1 X_0$, \\ $Y_3 X_2 X_1 Y_0$, $Y_3 X_2 Y_1 X_0$, $Y_3 Y_2 X_1 X_0$, $Y_3 Y_2 Y_1 Y_0$ \end{tabular}  & \begin{tabular}{c} $I_3 I_2 I_1 I_0$,   $I_3 I_2 Z_1 Z_0$,   $I_3 I_2 I_1 X_0$,   $I_3 I_2 \text{SWAP}_{(1-0)}$, \\  $I_3 \text{CS}_{(2-0)}$,   $I_3 Z_2 I_1 Z_0$,   $I_3 Z_2 Z_1 I_0$,   $\text{CS}^{(2)}_{(3-0)}$,\\   $Z_3 I_2 I_1 Z_0$,    $Z_3 I_2 Z_1 I_0$,   $Z_3 Z_2 I_1 I_0$,   $Z_3 Z_2 Z_1 Z_0$ \end{tabular} \\
        \hline
        $n_{\text{terms}}$ & & & $\num{2}^n$ & $\num{2}^{n-1} + n$ \\
        \bottomrule
        \end{tabular}
        \label{tab1}                        
    \end{center}
    \label{tab:decompositions_cfr}
\end{table*}

\section{Results}
The VQLS algorithm and the multi-qubit gates decomposition were used to solve two small tridiagonal linear systems.

The problem is the QLSP with $A$ defined in \eqref{eq:tridiag_generic} and with $\alpha=2$ and $\beta=-1$. Since the focus is mainly on the decomposition of $A$, the matrix $B$ was taken directly as unitary. In particular, $B = H^{\otimes n}$, where $H$ is the Hadamard operator and $n=\log_2(N)$ is the number of qubits.

The VQLS algorithm requires also the choice of an ansatz, a cost function and an optimizer. For this preliminary analyses, the simplest possible ansatz was adopted, which consists of a single $R_Y(\alpha)$ parametrized gate for every qubit, where
\begin{equation}
    R_Y(\alpha) = 
    \begin{bmatrix}
        \cos{\frac{\alpha}{2}} & -\sin{\frac{\alpha}{2}} \\
        \sin{\frac{\alpha}{2}} & \cos{\frac{\alpha}{2}}
    \end{bmatrix}.
    \label{eq:ry_gate}
\end{equation}
Furthermore, the global form of the cost function \eqref{eq:cost_global_norm} was used in both its normalized and non-normalized version. The non-normalized case is simply obtained by multiplying \eqref{eq:cost_global_norm} by $\brakettwo{\psi}{\psi}$. Finally, the COBYLA algorithm \cite{Powell1994} was used to minimize the cost function.

A $2\times 2$ and a $4\times 4$ linear systems were considered. In both cases, the problem was solved both on a simulator and on quantum hardware. In particular the `qasm\_simulator' and `ibmq\_athens' quantum backend from IBM \cite{IBMQ} were used. The decomposition, ansatz, cost-function and Hadamard-Test routines were implemented through the IBM Qiskit library for quantum programming \cite{Qiskit}.

\begin{figure*}[!t]
    \centering
    \subfloat[$2\times 2$]{\includegraphics[width=3in]{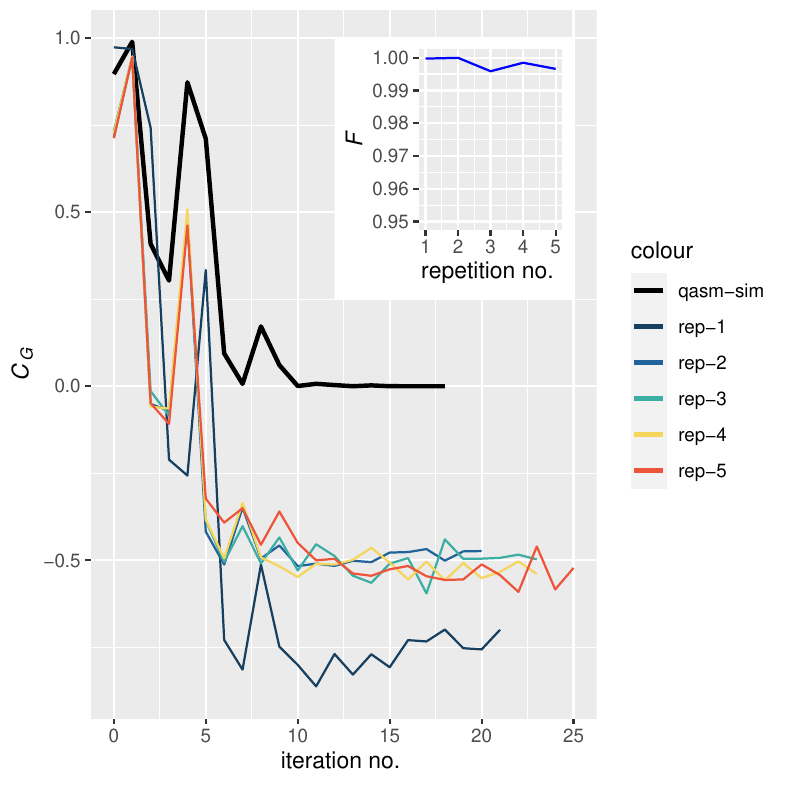}
    \label{fig:results_2b2}}
    \hfil
    \subfloat[$4\times 4$]{\includegraphics[width=3in]{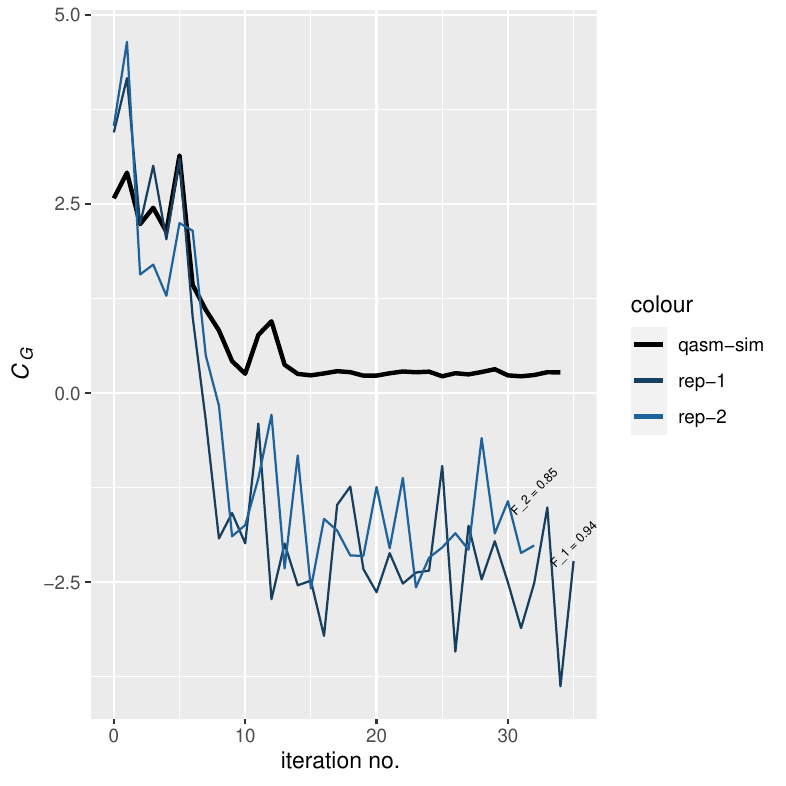}
    \label{fig:results_4b4}}
    \caption{Cost vs. number of iterations for two different tridiagonal quantum linear system problems (QLSPs). Both a simulator (`qasm\_simulator') and a real quantum computer (`ibmq\_athens') \cite{IBMQ} were used for the solution. On the left hand side is the $2\times 2$ case. While the cost assesses to 0 in the simulated case, it does not for the experiments on real hardware. This is due to undersampling and noise, which pollute the cost evaluation. Nevertheless, as shown by the inset, the runs on real hardware all deliver a final solution with high fidelity. On the right hand side is the $4\times 4$ case, in which the decomposition of $A$ includes the SWAP gate. This time, the cost function is non-normalized to improve convergence. Considerations similar to the $2\times 2$ problem apply here too. Even in this case, the cost at convergence differs between simulated and quantum runs. However the annotations next to the graphs show that high fidelities are reached at convergence.}
    \label{fig:results}
    \end{figure*}

The results from the $2\times 2$ linear system are represented in Fig. \ref{fig:results_2b2}. The cost function values for both the simulator and five quantum hardware runs are plotted. Also, the inset of Fig. \ref{fig:results_2b2} represents the fidelity of the solution at convergence. This is the scalar quantity
\begin{equation}
    F = |\brakettwo{x_{f,0}}{x_f}|^2,
    \label{eq:fidelity}
\end{equation}
where $x_{f,0}$ represents the analytical solution of \eqref{eq:QLSP}, while $x_f$ is the numerical solution obtained with VQLS. The fidelity $F$ is a measure of the overlapping between these two and it is thus 0 if the exact and numerical solutions are orthogonal and 1 if they are equal.

The plot shows that the cost vanishes for increasing iterations, when a simulator is used. However, the same does not hold for quantum hardware. Indeed, though the cost assess around the same final value for almost all runs, this value is not zero, but negative. This shows that the cost function was not estimated faithfully, which is likely due to two reasons. On the one hand, the number of samples was limited to 8192, generating an error due to undersampling. On the other hand, noise due to qubit relaxation and dephasing also affected the cost evaluation, once running on real hardware. Interestingly enough, however, despite the cost function being off by roughly 50\%, VQLS always assessed close to the correct final parameters, as demonstrated by the fidelity being very close to 1. This resilience of VQAs  to noise has been recently observed also in other applications \cite{Sharma2020}.

A similar analysis was performed also for the $4\times 4$ linear system and the results are shown in Fig. \ref{fig:results_4b4}. This case is more interesting, since this is the first dimension for which the Pauli and multi-qubit gates decompositions are not equal to each other. Differently than before, the cost function used was the non-normalized one, which was found to be more regular than \eqref{eq:cost_global_norm}. This explains the cost values higher than unity.

By looking at the simulator run, it can be observed that the cost does not assess to zero. This is likely due to the fact that the $R_Y$ ansatz cannot prefectly encode the exact solution. However, the best final fidelity in this case is a quite satisfactory 0.96. Regarding the runs executed on quantum hardware, a discrepancy is still observed due to undersampling and hardware noise. Also, the hardware availability so far allowed only to run two optimizations on ibmq\_athens, although more will follow in the near future. Even for the $4\times 4$ linear system, however, the noise in the cost function did not hinder the fidelity. In both runs in fact, the numerical solution found is close if not very close to the analytical one.

\bibliography{IEEEabrv,references}

\begin{thebibliography}{10}
\providecommand{\url}[1]{#1}
\csname url@samestyle\endcsname
\providecommand{\newblock}{\relax}
\providecommand{\bibinfo}[2]{#2}
\providecommand{\BIBentrySTDinterwordspacing}{\spaceskip=0pt\relax}
\providecommand{\BIBentryALTinterwordstretchfactor}{4}
\providecommand{\BIBentryALTinterwordspacing}{\spaceskip=\fontdimen2\font plus
\BIBentryALTinterwordstretchfactor\fontdimen3\font minus
  \fontdimen4\font\relax}
\providecommand{\BIBforeignlanguage}[2]{{%
\expandafter\ifx\csname l@#1\endcsname\relax
\typeout{** WARNING: IEEEtran.bst: No hyphenation pattern has been}%
\typeout{** loaded for the language `#1'. Using the pattern for}%
\typeout{** the default language instead.}%
\else
\language=\csname l@#1\endcsname
\fi
#2}}
\providecommand{\BIBdecl}{\relax}
\BIBdecl

\bibitem{Shor1999}
\BIBentryALTinterwordspacing
P.~W. Shor, ``Polynomial-time algorithms for prime factorization and discrete
  logarithms on a quantum computer,'' \emph{SIAM Review}, vol.~41, no.~2, pp.
  303--332, 1999. [Online]. Available:
  \url{http://www.jstor.org/stable/2653075}
\BIBentrySTDinterwordspacing

\bibitem{Grover1996}
\BIBentryALTinterwordspacing
L.~K. Grover, ``A fast quantum mechanical algorithm for database search,'' in
  \emph{Proceedings of the Twenty-Eighth Annual ACM Symposium on Theory of
  Computing}, ser. STOC '96.\hskip 1em plus 0.5em minus 0.4em\relax New York,
  NY, USA: Association for Computing Machinery, 1996, p. 212–219. [Online].
  Available: \url{https://doi.org/10.1145/237814.237866}
\BIBentrySTDinterwordspacing

\bibitem{Preskill2018}
J.~Preskill, ``Quantum {C}omputing in the {NISQ} era and beyond,''
  \emph{{Quantum}}, vol.~2, p.~79, Aug. 2018.

\bibitem{Arute2019}
\BIBentryALTinterwordspacing
F.~Arute \emph{et~al.}, ``Quantum supremacy using a programmable
  superconducting processor,'' \emph{Nature}, vol. 574, no. 7779, pp. 505--510,
  Oct. 2019. [Online]. Available:
  \url{https://doi.org/10.1038/s41586-019-1666-5}
\BIBentrySTDinterwordspacing

\bibitem{McClean2016}
\BIBentryALTinterwordspacing
J.~R. McClean, J.~Romero, R.~Babbush, and A.~Aspuru-Guzik, ``The theory of
  variational hybrid quantum-classical algorithms,'' \emph{New Journal of
  Physics}, vol.~18, no.~2, p. 023023, Feb. 2016. [Online]. Available:
  \url{https://doi.org/10.1088/1367-2630/18/2/023023}
\BIBentrySTDinterwordspacing

\bibitem{Peruzzo2014}
\BIBentryALTinterwordspacing
A.~Peruzzo, J.~McClean, P.~Shadbolt, M.-H. Yung, X.-Q. Zhou, P.~J. Love,
  A.~Aspuru-Guzik, and J.~L. O'Brien, ``A variational eigenvalue solver on a
  photonic quantum processor,'' \emph{Nature Communications}, vol.~5, no.~1,
  Jul. 2014. [Online]. Available: \url{https://doi.org/10.1038/ncomms5213}
\BIBentrySTDinterwordspacing

\bibitem{OMalley2016}
\BIBentryALTinterwordspacing
P.~O'Malley \emph{et~al.}, ``Scalable quantum simulation of molecular
  energies,'' \emph{Physical Review X}, vol.~6, no.~3, Jul. 2016. [Online].
  Available: \url{https://doi.org/10.1103/physrevx.6.031007}
\BIBentrySTDinterwordspacing

\bibitem{Jones2019}
\BIBentryALTinterwordspacing
T.~Jones, S.~Endo, S.~McArdle, X.~Yuan, and S.~C. Benjamin, ``Variational
  quantum algorithms for discovering hamiltonian spectra,'' \emph{Phys. Rev.
  A}, vol.~99, p. 062304, Jun 2019. [Online]. Available:
  \url{https://link.aps.org/doi/10.1103/PhysRevA.99.062304}
\BIBentrySTDinterwordspacing

\bibitem{Schuld2019}
\BIBentryALTinterwordspacing
M.~Schuld and N.~Killoran, ``Quantum machine learning in feature hilbert
  spaces,'' \emph{Phys. Rev. Lett.}, vol. 122, p. 040504, Feb 2019. [Online].
  Available: \url{https://link.aps.org/doi/10.1103/PhysRevLett.122.040504}
\BIBentrySTDinterwordspacing

\bibitem{Chen2020}
S.~Y.-C. Chen, C.-H.~H. Yang, J.~Qi, P.-Y. Chen, X.~Ma, and H.-S. Goan,
  ``Variational quantum circuits for deep reinforcement learning,'' \emph{IEEE
  Access}, vol.~8, pp. 141\,007--141\,024, 2020.

\bibitem{Romero2020}
\BIBentryALTinterwordspacing
J.~Romero and A.~Aspuru-Guzik, ``Variational quantum generators: Generative
  adversarial quantum machine learning for continuous distributions,''
  \emph{Advanced Quantum Technologies}, vol.~4, no.~1, p. 2000003, Dec. 2020.
  [Online]. Available: \url{https://doi.org/10.1002/qute.202000003}
\BIBentrySTDinterwordspacing

\bibitem{huang2019nearterm}
\BIBentryALTinterwordspacing
H.-Y. Huang, K.~Bharti, and P.~Rebentrost, ``Near-term quantum algorithms for
  linear systems of equations,'' 2019. [Online]. Available:
  \url{arXiv:1909.07344}
\BIBentrySTDinterwordspacing

\bibitem{xu2019variational}
\BIBentryALTinterwordspacing
X.~Xu, J.~Sun, S.~Endo, Y.~Li, S.~C. Benjamin, and X.~Yuan, ``Variational
  algorithms for linear algebra,'' 2019. [Online]. Available:
  \url{arXiv:1909.03898}
\BIBentrySTDinterwordspacing

\bibitem{Bravoprieto2020variational}
\BIBentryALTinterwordspacing
C.~Bravo-Prieto, R.~LaRose, M.~Cerezo, Y.~Subasi, L.~Cincio, and P.~J. Coles,
  ``Variational quantum linear solver,'' 2020. [Online]. Available:
  \url{arXiv:1909.05820v2}
\BIBentrySTDinterwordspacing

\bibitem{An2020quantum}
\BIBentryALTinterwordspacing
D.~An and L.~Lin, ``Quantum linear system solver based on time-optimal
  adiabatic quantum computing and quantum approximate optimization algorithm,''
  2020. [Online]. Available: \url{arXiv:1909.05500}
\BIBentrySTDinterwordspacing

\bibitem{Wu2021}
\BIBentryALTinterwordspacing
B.~Wu, M.~Ray, L.~Zhao, X.~Sun, and P.~Rebentrost, ``Quantum-classical
  algorithms for skewed linear systems with an optimized hadamard test,''
  \emph{Physical Review A}, vol. 103, no.~4, Apr. 2021. [Online]. Available:
  \url{https://doi.org/10.1103/physreva.103.042422}
\BIBentrySTDinterwordspacing

\bibitem{Harrow2009}
\BIBentryALTinterwordspacing
A.~W. Harrow, A.~Hassidim, and S.~Lloyd, ``Quantum algorithm for linear systems
  of equations,'' \emph{Phys. Rev. Lett.}, vol. 103, p. 150502, Oct 2009.
  [Online]. Available:
  \url{https://link.aps.org/doi/10.1103/PhysRevLett.103.150502}
\BIBentrySTDinterwordspacing

\bibitem{Quarteroni2014}
\BIBentryALTinterwordspacing
A.~Quarteroni, \emph{Numerical Models for Differential Problems}.\hskip 1em
  plus 0.5em minus 0.4em\relax Springer Milan, 2014. [Online]. Available:
  \url{https://doi.org/10.1007/978-88-470-5522-3}
\BIBentrySTDinterwordspacing

\bibitem{Knill1998}
\BIBentryALTinterwordspacing
E.~Knill and R.~Laflamme, ``Power of one bit of quantum information,''
  \emph{Phys. Rev. Lett.}, vol.~81, pp. 5672--5675, Dec 1998. [Online].
  Available: \url{https://link.aps.org/doi/10.1103/PhysRevLett.81.5672}
\BIBentrySTDinterwordspacing

\bibitem{Morimae2017}
\BIBentryALTinterwordspacing
T.~Morimae, ``Hardness of classically sampling the one-clean-qubit model with
  constant total variation distance error,'' \emph{Phys. Rev. A}, vol.~96, p.
  040302, Oct 2017. [Online]. Available:
  \url{https://link.aps.org/doi/10.1103/PhysRevA.96.040302}
\BIBentrySTDinterwordspacing

\bibitem{Fuji2018}
\BIBentryALTinterwordspacing
K.~Fujii, H.~Kobayashi, T.~Morimae, H.~Nishimura, S.~Tamate, and S.~Tani,
  ``Impossibility of classically simulating one-clean-qubit model with
  multiplicative error,'' \emph{Phys. Rev. Lett.}, vol. 120, p. 200502, May
  2018. [Online]. Available:
  \url{https://link.aps.org/doi/10.1103/PhysRevLett.120.200502}
\BIBentrySTDinterwordspacing

\bibitem{Kandala2017}
\BIBentryALTinterwordspacing
A.~Kandala, A.~Mezzacapo, K.~Temme, M.~Takita, M.~Brink, J.~M. Chow, and J.~M.
  Gambetta, ``Hardware-efficient variational quantum eigensolver for small
  molecules and quantum magnets,'' \emph{Nature}, vol. 549, no. 7671, pp.
  242--246, Sep. 2017. [Online]. Available:
  \url{https://doi.org/10.1038/nature23879}
\BIBentrySTDinterwordspacing

\bibitem{Hadfield2019}
\BIBentryALTinterwordspacing
S.~Hadfield, Z.~Wang, B.~O{\textquotesingle}Gorman, E.~Rieffel, D.~Venturelli,
  and R.~Biswas, ``From the quantum approximate optimization algorithm to a
  quantum alternating operator ansatz,'' \emph{Algorithms}, vol.~12, no.~2,
  p.~34, Feb. 2019. [Online]. Available:
  \url{https://doi.org/10.3390/a12020034}
\BIBentrySTDinterwordspacing

\bibitem{lloyd2018quantum}
\BIBentryALTinterwordspacing
S.~Lloyd, ``Quantum approximate optimization is computationally universal,''
  2018. [Online]. Available: \url{arXiv:1812.11075}
\BIBentrySTDinterwordspacing

\bibitem{Sweke2020}
\BIBentryALTinterwordspacing
R.~Sweke, F.~Wilde, J.~Meyer, M.~Schuld, P.~K. Faehrmann, B.~Meynard-Piganeau,
  and J.~Eisert, ``Stochastic gradient descent for hybrid quantum-classical
  optimization,'' \emph{Quantum}, vol.~4, p. 314, Aug. 2020. [Online].
  Available: \url{https://doi.org/10.22331/q-2020-08-31-314}
\BIBentrySTDinterwordspacing

\bibitem{Fletcher1987}
R.~Fletcher, \emph{Practical Methods of Optimization; (2nd Ed.)}.\hskip 1em
  plus 0.5em minus 0.4em\relax USA: Wiley-Interscience, 1987.

\bibitem{Nelder1965}
\BIBentryALTinterwordspacing
J.~A. Nelder and R.~Mead, ``A simplex method for function minimization,''
  \emph{The Computer Journal}, vol.~7, no.~4, pp. 308--313, Jan. 1965.
  [Online]. Available: \url{https://doi.org/10.1093/comjnl/7.4.308}
\BIBentrySTDinterwordspacing

\bibitem{Powell1994}
\BIBentryALTinterwordspacing
M.~J.~D. Powell, \emph{A Direct Search Optimization Method That Models the
  Objective and Constraint Functions by Linear Interpolation}.\hskip 1em plus
  0.5em minus 0.4em\relax Dordrecht: Springer Netherlands, 1994, pp. 51--67.
  [Online]. Available: \url{https://doi.org/10.1007/978-94-015-8330-5\_4}
\BIBentrySTDinterwordspacing

\bibitem{Nielsen2010}
M.~A. Nielsen and I.~L. Chuang, \emph{Quantum Computation and Quantum
  Information: 10th Anniversary Edition}.\hskip 1em plus 0.5em minus
  0.4em\relax Cambridge University Press, 2010.

\bibitem{IEEEexample:Cappanera2021}
E.~Cappanera, ``A variational linear solver for the poisson 1d matrix,''
  Master's thesis, 2021, to be published.

\bibitem{Qiskit}
H.~Abraham \emph{et~al.}, ``Qiskit: An open-source framework for quantum
  computing,'' 2019.

\bibitem{IBMQ}
\BIBentryALTinterwordspacing
``Ibm quantum,'' 2021. [Online]. Available:
  \url{https://quantum-computing.ibm.com/}
\BIBentrySTDinterwordspacing

\bibitem{Sharma2020}
\BIBentryALTinterwordspacing
K.~Sharma, S.~Khatri, M.~Cerezo, and P.~J. Coles, ``Noise resilience of
  variational quantum compiling,'' \emph{New Journal of Physics}, vol.~22,
  no.~4, p. 043006, Apr. 2020. [Online]. Available:
  \url{https://doi.org/10.1088/1367-2630/ab784c}
\BIBentrySTDinterwordspacing

\end{thebibliography}

\end{document}